\documentclass[twocolumn,showpacs,preprintnumbers,amsmath,amssymb,superscriptaddress,prb]{revtex4}
\usepackage{graphicx}
\begin{document}

\title{Dilute moment n-type ferromagnetic semiconductor Li(Zn,Mn)As}

\author{J. Ma\v{s}ek}
\author{J.Kudrnovsk\'y}
\author{F. M\'aca}
\affiliation{Institute of Physics ASCR, Na Slovance 2, 182 21 Praha 8, Czech Republic}

\author{B. L. Gallagher}
\author{R. P. Campion}
\affiliation{School of Physics and Astronomy, University of Nottingham, Nottingham NG7 2RD, UK}

\author{D. H. Gregory}
\affiliation{Department of Chemistry, University of Glasgow, Glasgow G12 8QQ, UK}

\author{T. Jungwirth}
\affiliation{Institute of Physics ASCR, Cukrovarnick\'a 10, 162 53
Praha 6, Czech Republic}
\affiliation{School of Physics and Astronomy, University of Nottingham, Nottingham NG7 2RD, UK}

\begin{abstract}
We propose to replace Ga in (Ga,Mn)As with Li and Zn as a route to high Curie temperature, carrier mediated
ferromagnetism in a dilute moment n-type semiconductor. 
Superior material characteristics, rendering Li(Zn,Mn)As a realistic candidate for such a system, include 
unlimited solubility of the isovalent substitutional Mn
impurity and carrier concentration controlled 
independently of Mn doping by adjusting Li-(Zn,Mn) stoichiometry.
Our predictions
are anchored by detail {\em ab initio} calculations and comparisons with the familiar and directly related
(Ga,Mn)As, by the microscopic physical picture we provide for the
exchange interaction
between Mn local moments and electrons in the conduction band, and by analysis of  
prospects for the controlled growth of Li(Zn,Mn)As materials. 
\end{abstract}
\maketitle

(Ga,Mn)As is a prototypical of a unique class of spintronic materials 
in which ferromagnetic
coupling between dilute local moments is mediated by semiconductor band 
carriers.\cite{Dietl:2000_a,Jungwirth:2006_a} This unusual  
behavior is realized in  the conventional semiconductor GaAs using only one type of dopant,  Mn$_{\rm Ga}$,
which
provides both local spins and holes. The simplicity of such a 
system inevitably brings also  limitations to the structural, ferromagnetic and semiconducting
properties due to the low solubility of Mn and due to the lack of independent
control of local moment and carrier densities. Among the negative consequences 
are Curie temperatures
below room temperature, and p-type conduction only.  
In our study we predict that  these limitations can be lifted by 
a straightforward substitution of
the group-III element Ga with group-I Li and group-II Zn elements.

\begin{figure}[h]

\vspace*{-1cm}
\hspace*{.8cm}\includegraphics[width=.3\textwidth,angle=-90]{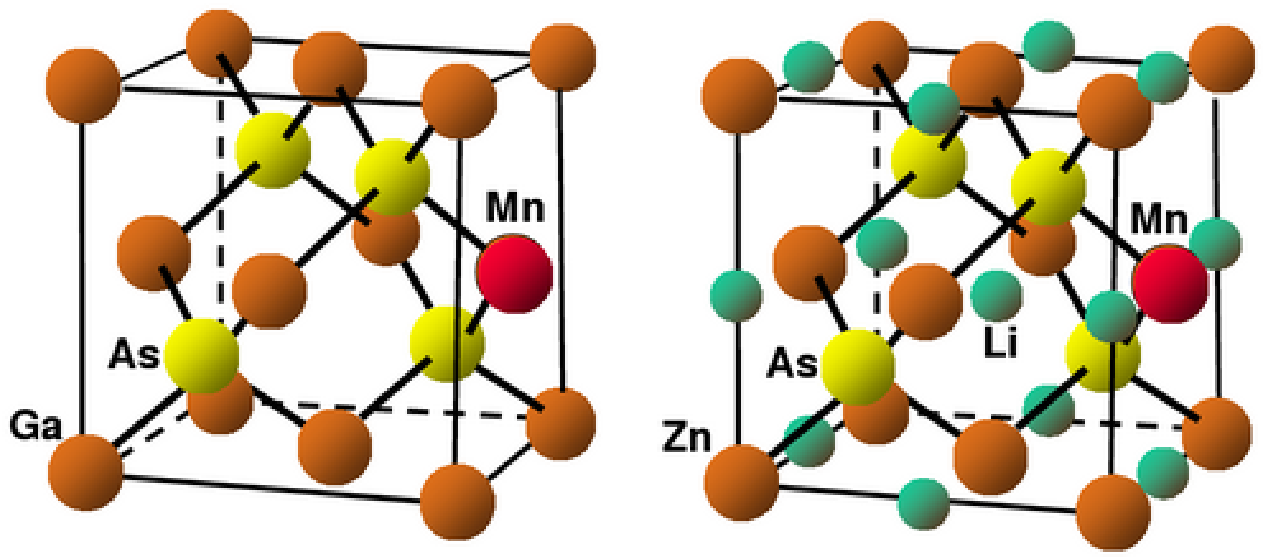}\hspace{0.5cm}

\vspace*{-1.5cm}
\hspace*{0cm}\includegraphics[width=.3\textwidth,angle=-90]{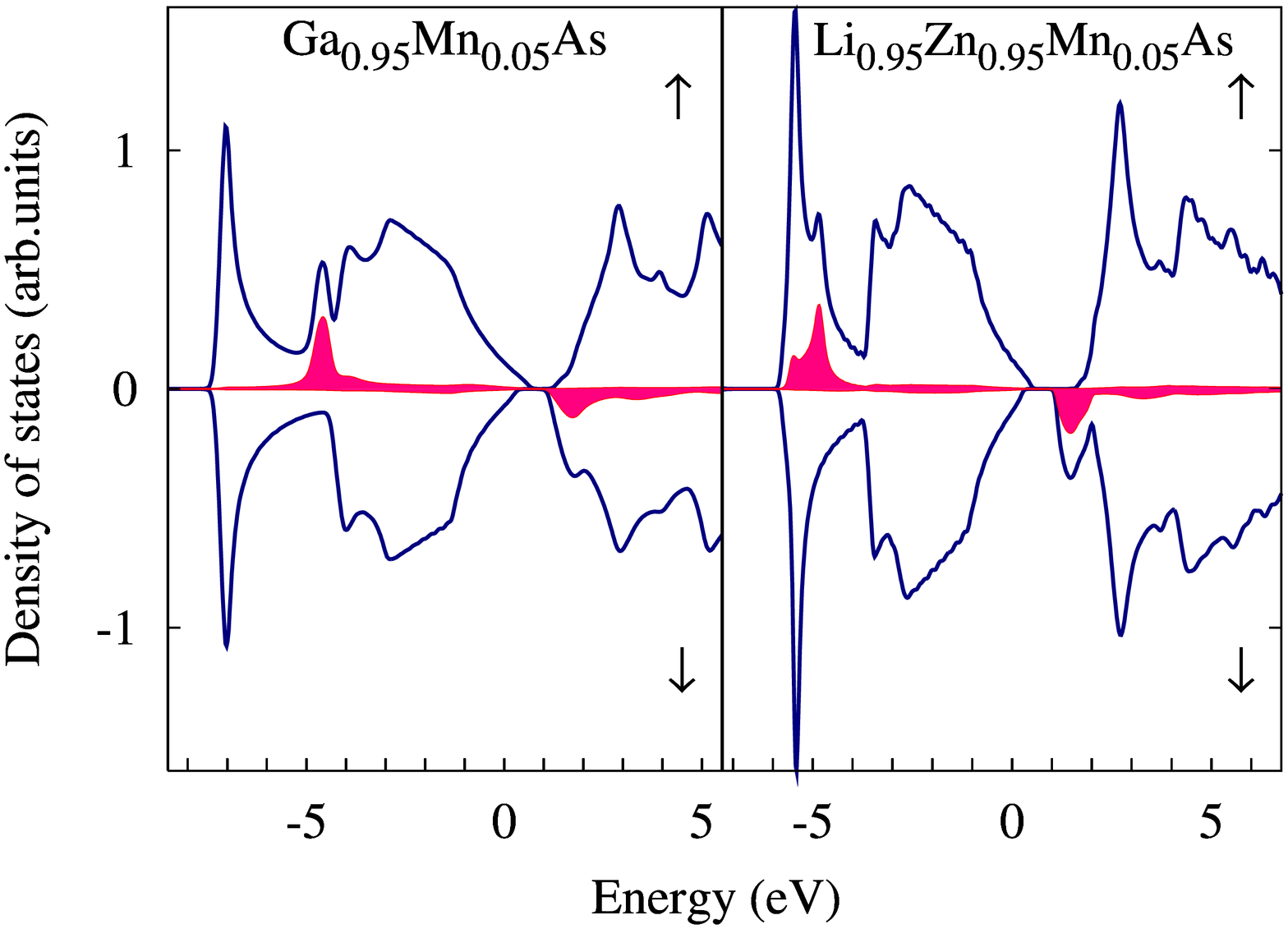}

\caption{Top panel: Schematics of (Ga,Mn)As and Li(Zn,Mn)As crystal structures. Bottom panel: 
{\em ab initio} total (blue lines) and Mn $d$-orbital projected (red filled lines) density 
of states  of Ga$_{0.95}$Mn$_{0.05}$As and Li$_{0.95}$(Zn$_{0.95}$Mn$_{0.05}$)As mixed crystals. Energy on the
$x$-axis is measured from the Fermi energy.}
\label{cryst}
\end{figure}

LiZnAs is a stable direct-gap semiconductor which
can be grown by  the
high temperature
reaction of elemental Li, Zn, and As.\cite{Bacewicz:1988_a,Kuriyama:1987_a} 
Its crystal structure and band structure are very 
similar to those of GaAs. As shown in  Fig.~\ref{cryst}(a) and (b), the
LiZnAs tetrahedral lattice can be viewed as a zincblende ZnAs binary compound, analogous to GaAs,
filled with Li atoms  at  tetrahedral interstitial
sites near As.  The relatively small first ionization
energy of  Li makes the (Li)$^{+}$(ZnAs)$^{-}$ 
compound half ionic and half covalent.\cite{Wei:1986_a} 
The measured band gap
of LiZnAs (1.61~eV) is very similar to the GaAs band-gap (1.52~eV).\cite{Kuriyama:1994_a} An overall similarity
of the ${\rm LiZnAs}$ and GaAs electronic structures, including valence band and conduction band dispersions, 
ground state charge density, and phonon dispersion relations has
been  reported in  {\em ab initio} local-density-approximation (LDA)   studies.\cite{Wei:1986_a,Wood:2005_a}  

Application of the LDA description of the electronic band-structure
to Mn-doped zincblende semiconductors
runs into a conceptual difficulty in dealing with local moment levels coincident
with itinerant electron bands. To partly remedy this inadequacy of standard first principles approaches
we use in this paper the LDA+U technique which combines LDA with the Hubbard description of strongly
correlated localized orbitals.\cite{Anisimov:1991_a,Park:2000_a,Shick:2004_a,Wierzbowska:2004_a} 
The LDA+U method is implemented within the framework of the first-principles, tight-binding linear muffin-tin orbital 
approach;
disorder effects  associated with random distribution of Mn and other defects if present 
are accounted for using the coherent potential
approximation (CPA).\cite{Kudrnovsky:2004_a}

In Fig.~\ref{cryst}(c) and (d) we compare LDA+U/CPA  density of states (DOS)  of LiZnAs and GaAs doped with
5\% of Mn.  Since Mn introduces one hole per Mn in GaAs but is isovalent in LiZnAs, the structure considered
in Fig.~\ref{cryst}(d) is Li$_{0.95}$(Zn$_{0.95}$Mn$_{0.05}$)As to allow for a direct comparison of p-type
ferromagnets realized in the two hosts. As expected from the similar host band structures and nearly
atomic-like character of Mn $d$-states in (moderately) narrow-gap zincblende semiconductors,\cite{Schulthess:2005_a} 
both systems are
ferromagnetic with comparable valence-band exchange splittings and similar Mn $d$ and host $sp$-projected DOSs.  In particular, the main peak in the
majority-spin $d$-orbital DOS is well below the valence band edge resulting in states
near the Fermi energy having a predominantly As $p$-character
in both systems. 

The LDA+U/CPA band structure of (Ga,Mn)As is consistent with a number of experiments
in this ferromagnetic semiconductor.\cite{Jungwirth:2006_a} We expect, 
based on the above similarities between the two systems, that Li(Zn,Mn)As is also described reliably
by this theoretical technique. In the following paragraphs we use
the {\em ab initio} band structures to explore Li(Zn,Mn)As mixed crystals
over a wide range of local moment and carrier dopings, including regimes 
which are technologically inaccessible
in the GaAs host. 
In particular, we focus on the equilibrium solubility 
of isovalent Mn in LiZnAs for concentrations
well above 1\% and on ferromagnetism in LiZnAs with
n-type conduction which, as we show, can be readily achieved without introducing any additional chemical elements.

\begin{figure}[h]
\vspace*{-0.5cm}

\includegraphics[width=.31\textwidth,angle=-90]{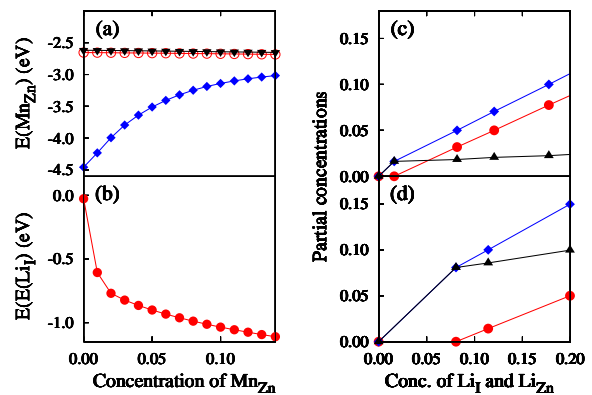}
\caption{{\em Ab initio} impurity formation energies of Li(Zn,Mn)As mixed crystals: (a) Mn$_{\rm Zn}$
formation energy as a function of Mn$_{\rm Zn}$ doping 
for  stoichiometric structures (red open dots), for p-type systems with 5\% of Li vacancy acceptors 
(black triangles),
and for n-type systems with 5\% of donor Li$_{\rm I}$ impurities occupying Zn-tetrahedral interstitial sites 
(blue diamonds).
(b) Li$_{\rm I}$ formation energy
near the stoichiometric composition as a function of Mn$_{\rm Zn}$ doping. (c) Partial concentration of Li$_{\rm I}$
(blue diamonds) donors, Li$_{\rm Zn}$ acceptors (red dots), and net electron doping (black triangles) for 5\% of 
Mn$_{\rm Zn}$ as a function of total concentration of excess Li (Li$_{\rm I}$ plus Li$_{\rm Zn}$).
(d) Same as (c) for 12.5\% of Mn$_{\rm Zn}$.}
\label{part}
\end{figure}

The calculated formation energies of Mn$_{\rm Zn}$ are plotted in Fig.~\ref{part}(a) for insulating 
stoichiometric Li(Zn,Mn)As,
for p-type  Li(Zn,Mn)As  with Li vacancies, and for n-type  Li(Zn,Mn)As with additional Li$_{\rm I}$ 
atoms occupying tetrahedral
interstitial sites near Zn. Independent of the charge doping, the formation energy of Mn$_{\rm Zn}$ is 
negative, i.e., we find no equilibrium solubility limit for substitutional Mn$_{\rm Zn}$, consistent
with its isovalent nature. (Note that we focus
in this paper on dilute Mn systems far from  the LiMnAs limit; the LiMnAs compound  is an  
antiferromagnet\cite{Achenbach:1981_a,Bronger:1986_a} due to short-range Mn-Mn superexchange.)
This result contrasts with the positive
formation energy of Mn$_{\rm Ga}$ in GaAs\cite{Masek:2004_a} which leads to an 
equilibrium solubility limit below 1\%. 

A detailed analysis of non-stoichiometric n-type Li(Zn,Mn)As is presented in Figs.~\ref{part}(b)-(d).
Panel (b) demonstrates that in systems with over-stoichiometric Li concentrations a large number of 
Li$_{\rm I}$ atoms can be
incorporated at interstitial sites near Zn. Formation energy of these single-donor impurities is negative and 
decreases with increasing Mn doping. 

In covalent semiconductors,  carrier doping is limited
due to strong tendency to self-compensation. 
In the ${\rm LiZnAs}$ host with excess Li, Li$_{\rm Zn}$ antisites represent natural single-acceptor defects compensating the interstitial
Li$_{\rm I}$ donors. {\em Ab initio} calculations of formation energies can be 
used\cite{Masek:2004_a,Jungwirth:2005_b} to estimate
the dependence of  Li$_{\rm I}$ and  Li$_{\rm Zn}$ partial concentrations
on  the total density of excess Li atoms above the Li-(Zn,Mn) stoichiometry. Results are plotted in 
Figs.~\ref{part}(c) and (d) for Mn doping of 5\% and 12.5\%, respectively. Although the tendency to self-compensation
is clearly apparent, the theoretical data  suggest  that large net electron densities are feasible in Li(Zn,Mn)As owing
to
the partly ionic character of the compound. Remarkably, the n-type doping efficiency by excess Li increases significantly 
with increasing Mn concentration.

\begin{figure}[h]
\vspace*{-0.5cm}

\includegraphics[width=.45\textwidth,angle=-0]{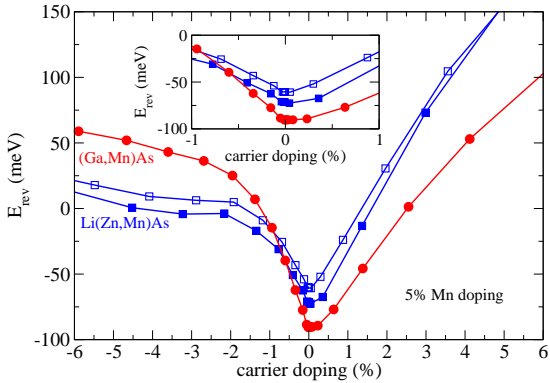}
\vspace*{-0.5cm}
\caption{Energy
cost of flipping one Mn moment calculated 
with all other Mn moments aligned. 
Open and closed squares  were obtained from n-type
Li$_{1.05}$(Zn$_{0.95}$,Mn$_{0.05}$)As and p-type Li$_{0.95}$(Zn$_{0.95}$,Mn$_{0.05}$)As {\em ab initio} band
structures, respectively (see text for details). Circles represent calculations for  Ga$_{0.95}$Mn$_{0.05}$As. 
}
\label{tc}
\end{figure}

We now consider the magnetic
properties of Li(Zn,Mn)As semiconductors. The
compatibility of the CPA with Weiss mean-field theory allows us to estimate the
strength of Mn-Mn magnetic coupling,  at a given chemical composition, from the calculated energy
cost, $E_{rev}$, of reversing one Mn moment with all other Mn moments aligned.\cite{Liechtenstein:1987_a,Masek:1991_a}
Results for  $E_{rev}$ shown in Fig.~\ref{tc}(a) were obtained using a rigid-band approximation. Here the LDA+U/CPA
band structure was calculated in one charge state of the compound only, and the dependence on carrier doping was obtained
by shifting the Fermi energy while keeping the bands fixed. Note that for the 5\% Mn doping considered in the figure, 
carrier
doping of $-5$\% corresponds to one hole per Mn and $+5$\% to one electron per Mn.
Open and closed squares  represent  $E_{rev}$ obtained applying the above rigid band scheme, based on n-type
Li$_{1.05}$(Zn$_{0.95}$,Mn$_{0.05}$)As and p-type Li$_{0.95}$(Zn$_{0.95}$,Mn$_{0.05}$)As band
structures, respectively. 
The similarity of the two curves indicates that the position of the Fermi energy plays the dominant role in magnetic
interactions while disorder effects associated 
with non-stoichiometric Li-(Zn,Mn) configurations are less important. 
For comparison we plot in Fig.~\ref{tc}(a) the rigid band $E_{rev}$ obtained from  the
Ga$_{0.95}$Mn$_{0.05}$As {\em ab initio} band structure (see also Ref.~\onlinecite{Sandratskii:2002_a}). As expected
the two compounds show very similar magnetic behavior. 

\begin{figure}[h]
\vspace*{-0.5cm}

\includegraphics[width=.32\textwidth,angle=-90]{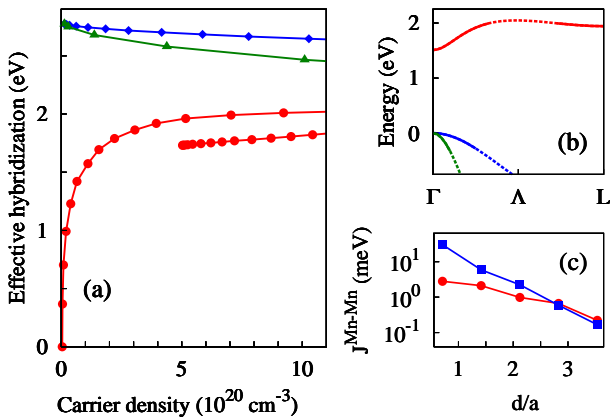}
\caption{(a) Model calculations of Mn$_{\rm Ga}$ $d$-orbital hybridization potentials with
GaAs $sp$-orbitals at the Fermi energy in the valence band (blue diamonds and green triangles) 
and conduction band (red dots) as a 
function of hole density and electron density, respectively. 
(b) Model band structure of the GaAs host illustrating the filled conduction
band states (red full line) and empty valence band states ( black and blue lines)
for electron or hole density of  10$^{21}$ cm$^{-3}$ (largest carrier density
considered in panel (a)).
(c) {\em Ab initio} Mn-Mn exchange interactions as a function of Mn interatomic
distance evaluated along the [110] crystal direction. 
}
\label{interp}
\end{figure}

The calculated positive values of $E_{rev}$ at large hole concentrations
correspond to ferromagnetic Mn-Mn coupling, which is consistent with experimentally
observed ferromagnetic ground states in highly doped p-type (Ga,Mn)As. 
(At low enough hole
concentrations, $E_{rev}$ changes sign and ferromagnetism becomes unstable due to the dominant role
of short-range antiferromagnetic Mn-Mn superexchange.\cite{Jungwirth:2006_a})
The comparable
strength of ferromagnetic Mn-Mn coupling for n-type and p-type doping at high carrier concentrations 
suggests that exchange interactions
between Mn $d$-states and conduction band states in zincblende
semiconductors can be strong and that  Li(Zn,Mn)As with properly adjusted
stoichiometry is a high Curie temperature n-type ferromagnetic semiconductor. 

We now discuss the physical origin of the high temperature electron-mediated ferromagnetism.
In zincblende semiconductors doped with Mn $d^5$ impurities, the local moments can have either a  
direct exchange interaction with band electrons on the same site and/or an interaction due to $sp-d$ 
hybridization between the Mn local moment and band electrons on neighboring 
sites.\cite{Bhattacharjee:1983_a,Dietl:1994_a}  The latter interaction is much stronger and high-temperature ferromagnetism
is expected to occur only when the hybridization  coupling mechanism is present.
In GaAs,  the valence band edge at the $\Gamma$-point has a $p$-orbital character with a larger contribution from 
As.\cite{Harrison:1980_a}
Since the density of states at the valence band
edge is large, the occupied states for typical hole densities in ferromagnetic (Ga,Mn)As are still
close to the $\Gamma$-point and have  similar atomic orbital composition. The $p-d$ hybridization is allowed by symmetry within
the whole Fermi sea (including the  $\Gamma$-point) which, together with the large magnetic susceptibility of heavy 
 mass holes, explains the experimentally observed 
 robust ferromagnetism in p-type (Ga,Mn)As. 

The bottom of the conduction band of GaAs has an $s$-orbital character
with a larger contribution from 
Ga.  The smaller admixture of orbitals from the nearest neighbor As
atoms
to Mn$_{\rm Ga}$, the vanishing $s-d$ hybridization at the $\Gamma$-point due to symmetry, and  the small magnetic susceptibility of  low effective mass electrons 
near the bottom of the conduction band all suggest that ferromagnetism is unfavorable in  n-type 
systems. As shown in the inset of Fig.~\ref{tc}, a weaker tendency towards ferromagnetism in n-type materials
is indeed observed in our {\em ab initio} calculations, however,
is limited to  relatively low electron dopings.

To explain the seemingly unexpected robust ferromagnetism at larger electron concentrations
we present, in Fig.~\ref{interp}(a) and (b),  tight-binding model
calculations of the energy dependent $sp-d$ hybridization potential in
the conduction band and compare with the more familiar valence band case. 
When moving off the $\Gamma$-point, the hybridization rapidly sets in partly due to
the $s-d$ hybridization allowed by symmetry at non-zero wavevectors and partly due to 
the admixture of As and Ga $p$-orbitals
which increases with increasing wavevector. 
The strength of the $sp-d$ hybridization
in highly doped n-type systems becomes comparable to that of p-type materials as the Fermi energy 
leaves the bottom part of the conduction
band and approaches the $L$-point.
The same picture, which we have discussed for the more familiar GaAs host doped with Mn, applies to 
Li(Zn,Mn)As whose
material properties, as analyzed above,  are much more favorable for achieving large n-type doping.
 
Within the mean-field theory, the ferromagnetic
Curie temperature, $T_c$, can be estimated as $T_c\approx E_{rev}/6k_B$, where  
$k_B$ is the Boltzmann constant. The mean-field approximation should be
reliable when the carrier
mediated Mn-Mn coupling is sufficiently long range  
but tends to overestimate $T_c$ when the carriers become more localized and magnetic interactions
short-ranged.\cite{Sato:2004_a,Bergqvist:2004_a,Bouzerar:2004_a}
In Fig.~\ref{interp}(c) we plot the spatially dependent 
interatomic exchange energies obtained by mapping the LDA+U/CPA total energies  
on the Heisenberg Hamiltonian.\cite{Kudrnovsky:2004_a}
As seen in the figure, the n-type Li(Zn,Mn)As shows very similar magnetic interaction characteristics
to the p-type (Ga,Mn)As with  corresponding doping densities. In both systems the leading exchange interactions
are ferromagnetic and the interaction range safely exceeds the average Mn-Mn moment distance.
Consistent with the theory expectations, experimental high quality (Ga,Mn)As materials show mean field-like
magnetization curves and Curie temperatures  proportional to the density of Mn$_{\rm Ga}$.\cite{Jungwirth:2005_b} 
Maximum
Mn$_{\rm Ga}$ doping achieved so far is approximately 6\% and the corresponding record $T_c$ is 
173~K.\cite{Jungwirth:2005_b}  Calculations shown in  Fig.~\ref{interp} suggest comparable Curie temperatures
for the n-type Li(Zn,Mn)As counterparts. Since we found 
no doping limit for  Mn$_{\rm Zn}$ in Li(Zn,Mn)As and straightforward means of high electron doping in the material,
Li(Zn,Mn)As might lead not only
to the realization of a high Curie temperature diluted magnetic semiconductor with electron conduction
but also might allow for larger magnetic moment density and therefore larger maximum $T_c$ than (Ga,Mn)As.

We conclude by briefly discussing prospects for the controlled growth of
Li(Zn,Mn)As materials. Bulk high
structural quality LiZnAs crystallites have been formed by reaction of
near equimolar elemental Li, Zn, and As.\cite{Bacewicz:1988_a,Kuriyama:1987_a} The crystallites are stable in
inert atmospheres but slowly oxidize in air. It is probable that Li(Zn,Mn)As
crystallites could be formed in the same way. However, to obtain the
required control of stoichiometry to produce the ferromagnetic
semiconductors predicted by our calculations, molecular beam epitaxy (MBE)
is likely required. LiZnAs has not previously been grown by MBE, presumably
because non-magnetic LiZnAs does not seem to have any advantage over
GaAs. The similarity of Li(Zn,Mn)As to (Ga,Mn)As suggests that it should be
possible to grow Li(Zn,Mn)As by MBE;
here Li can be supplied from a
metal-organic gas source. Li(Zn,Mn)As could be grown by MBE on lattice
matched, relaxed, (In,Ga)As epilayers on normal GaAs substrates, carrier
doping achieved by careful control of the Li flux, and, furthermore, it
should be possible to avoid post growth oxidation by overgrowth of the
Li(Zn,Mn)As epilayers.

We acknowledge discussions with  C.T. Foxon, A.H. MacDonald,  
J. Sinova, and support from the
Grant Agency of the Czech
Republic under Grant 202/05/0575 and  202/04/0583, the Academy of Sciences of the
Czech Republic under Institutional Support AVOZ10100521 and AVOZ10100520, 
the Ministry of Education of the
Czech Republic Center for Fundamental Research LC510 and COST P19 OC-150, the 
EPSRC under Grant GR/S81407/01, and the National Science
Foundation under Grant No. PHY99-07949.


\begin{thebibliography}{27}
\expandafter\ifx\csname natexlab\endcsname\relax\def\natexlab#1{#1}\fi
\expandafter\ifx\csname bibnamefont\endcsname\relax
  \def\bibnamefont#1{#1}\fi
\expandafter\ifx\csname bibfnamefont\endcsname\relax
  \def\bibfnamefont#1{#1}\fi
\expandafter\ifx\csname citenamefont\endcsname\relax
  \def\citenamefont#1{#1}\fi
\expandafter\ifx\csname url\endcsname\relax
  \def\url#1{\texttt{#1}}\fi
\expandafter\ifx\csname urlprefix\endcsname\relax\def\urlprefix{URL }\fi
\providecommand{\bibinfo}[2]{#2}
\providecommand{\eprint}[2][]{\url{#2}}

\bibitem[{\citenamefont{Dietl et~al.}(2000)\citenamefont{Dietl, Ohno,
  Matsukura, Cibert, and Ferrand}}]{Dietl:2000_a}
\bibinfo{author}{\bibfnamefont{T.}~\bibnamefont{Dietl}},
  \bibinfo{author}{\bibfnamefont{H.}~\bibnamefont{Ohno}},
  \bibinfo{author}{\bibfnamefont{F.}~\bibnamefont{Matsukura}},
  \bibinfo{author}{\bibfnamefont{J.}~\bibnamefont{Cibert}}, \bibnamefont{and}
  \bibinfo{author}{\bibfnamefont{D.}~\bibnamefont{Ferrand}},
  \bibinfo{journal}{Science} \textbf{\bibinfo{volume}{287}},
  \bibinfo{pages}{1019} (\bibinfo{year}{2000}).

\bibitem[{\citenamefont{Jungwirth et~al.}(2006)\citenamefont{Jungwirth, Sinova,
  {Ma\v{s}ek}, {Ku\v{c}era}, and MacDonald}}]{Jungwirth:2006_a}
\bibinfo{author}{\bibfnamefont{T.}~\bibnamefont{Jungwirth}},
  \bibinfo{author}{\bibfnamefont{J.}~\bibnamefont{Sinova}},
  \bibinfo{author}{\bibfnamefont{J.}~\bibnamefont{{Ma\v{s}ek}}},
  \bibinfo{author}{\bibfnamefont{J.}~\bibnamefont{{Ku\v{c}era}}},
  \bibnamefont{and} \bibinfo{author}{\bibfnamefont{A.~H.}
  \bibnamefont{MacDonald}}, \bibinfo{journal}{Rev. Mod. Phys.}
  \textbf{\bibinfo{volume}{78}}, \bibinfo{pages}{809} (\bibinfo{year}{2006}).

\bibitem[{\citenamefont{Wei and Zunger}(1986)}]{Wei:1986_a}
\bibinfo{author}{\bibfnamefont{S.-H.} \bibnamefont{Wei}} \bibnamefont{and}
  \bibinfo{author}{\bibfnamefont{A.}~\bibnamefont{Zunger}},
  \bibinfo{journal}{Phys. Rev. Lett.} \textbf{\bibinfo{volume}{56}},
  \bibinfo{pages}{528} (\bibinfo{year}{1986}).

\bibitem[{\citenamefont{Bacewicz and Ciszek}(1988)}]{Bacewicz:1988_a}
\bibinfo{author}{\bibfnamefont{R.}~\bibnamefont{Bacewicz}} \bibnamefont{and}
  \bibinfo{author}{\bibfnamefont{T.~F.} \bibnamefont{Ciszek}},
  \bibinfo{journal}{Appl. Phys. Lett.} \textbf{\bibinfo{volume}{52}},
  \bibinfo{pages}{1150} (\bibinfo{year}{1988}).

\bibitem[{\citenamefont{Kuriyama et~al.}(1994)\citenamefont{Kuriyama, Kato, and
  Kawada}}]{Kuriyama:1994_a}
\bibinfo{author}{\bibfnamefont{K.}~\bibnamefont{Kuriyama}},
  \bibinfo{author}{\bibfnamefont{T.}~\bibnamefont{Kato}}, \bibnamefont{and}
  \bibinfo{author}{\bibfnamefont{K.}~\bibnamefont{Kawada}},
  \bibinfo{journal}{Phys. Rev.} \textbf{\bibinfo{volume}{B 49}},
  \bibinfo{pages}{11452} (\bibinfo{year}{1994}).

\bibitem[{\citenamefont{Wood and Strohmayer}(2005)}]{Wood:2005_a}
\bibinfo{author}{\bibfnamefont{D.~M.} \bibnamefont{Wood}} \bibnamefont{and}
  \bibinfo{author}{\bibfnamefont{W.~H.} \bibnamefont{Strohmayer}},
  \bibinfo{journal}{Phys. Rev.} \textbf{\bibinfo{volume}{B 71}},
  \bibinfo{pages}{193201} (\bibinfo{year}{2005}).

\bibitem[{\citenamefont{Kuriyama and Nakamura}(1987)}]{Kuriyama:1987_a}
\bibinfo{author}{\bibfnamefont{K.}~\bibnamefont{Kuriyama}} \bibnamefont{and}
  \bibinfo{author}{\bibfnamefont{F.}~\bibnamefont{Nakamura}},
  \bibinfo{journal}{Phys. Rev.} \textbf{\bibinfo{volume}{B 36}},
  \bibinfo{pages}{4439} (\bibinfo{year}{1987}).

\bibitem[{\citenamefont{Anisimov et~al.}(1991)\citenamefont{Anisimov, Zaanen,
  and Andersen}}]{Anisimov:1991_a}
\bibinfo{author}{\bibfnamefont{V.~I.} \bibnamefont{Anisimov}},
  \bibinfo{author}{\bibfnamefont{J.}~\bibnamefont{Zaanen}}, \bibnamefont{and}
  \bibinfo{author}{\bibfnamefont{O.~K.} \bibnamefont{Andersen}},
  \bibinfo{journal}{Phys. Rev.} \textbf{\bibinfo{volume}{B 44}},
  \bibinfo{pages}{943} (\bibinfo{year}{1991}).

\bibitem[{\citenamefont{Park et~al.}(2000)\citenamefont{Park, Kwon, and
  Min}}]{Park:2000_a}
\bibinfo{author}{\bibfnamefont{J.~H.} \bibnamefont{Park}},
  \bibinfo{author}{\bibfnamefont{S.~K.} \bibnamefont{Kwon}}, \bibnamefont{and}
  \bibinfo{author}{\bibfnamefont{B.~I.} \bibnamefont{Min}},
  \bibinfo{journal}{Physica} \textbf{\bibinfo{volume}{B 281/282}},
  \bibinfo{pages}{703} (\bibinfo{year}{2000}).

\bibitem[{\citenamefont{Shick et~al.}(2004)\citenamefont{Shick,
  {Kudrnovsk\'{y}}, and Drchal}}]{Shick:2004_a}
\bibinfo{author}{\bibfnamefont{A.~B.} \bibnamefont{Shick}},
  \bibinfo{author}{\bibfnamefont{J.}~\bibnamefont{{Kudrnovsk\'{y}}}},
  \bibnamefont{and} \bibinfo{author}{\bibfnamefont{V.}~\bibnamefont{Drchal}},
  \bibinfo{journal}{Phys. Rev.} \textbf{\bibinfo{volume}{B 69}},
  \bibinfo{pages}{125207} (\bibinfo{year}{2004}).

\bibitem[{\citenamefont{Wierzbowska et~al.}(2004)\citenamefont{Wierzbowska,
  Sanchez-Portal, and Sanvito}}]{Wierzbowska:2004_a}
\bibinfo{author}{\bibfnamefont{M.}~\bibnamefont{Wierzbowska}},
  \bibinfo{author}{\bibfnamefont{D.}~\bibnamefont{Sanchez-Portal}},
  \bibnamefont{and} \bibinfo{author}{\bibfnamefont{S.}~\bibnamefont{Sanvito}},
  \bibinfo{journal}{Phys. Rev.} \textbf{\bibinfo{volume}{B 70}},
  \bibinfo{pages}{235209} (\bibinfo{year}{2004}).

\bibitem[{\citenamefont{{Kudrnovsk\'{y}}
  et~al.}(2004)\citenamefont{{Kudrnovsk\'{y}}, Turek, Drchal, {M\'{a}ca},
  Weinberger, and Bruno}}]{Kudrnovsky:2004_a}
\bibinfo{author}{\bibfnamefont{J.}~\bibnamefont{{Kudrnovsk\'{y}}}},
  \bibinfo{author}{\bibfnamefont{I.}~\bibnamefont{Turek}},
  \bibinfo{author}{\bibfnamefont{V.}~\bibnamefont{Drchal}},
  \bibinfo{author}{\bibfnamefont{F.}~\bibnamefont{{M\'{a}ca}}},
  \bibinfo{author}{\bibfnamefont{P.}~\bibnamefont{Weinberger}},
  \bibnamefont{and} \bibinfo{author}{\bibfnamefont{P.}~\bibnamefont{Bruno}},
  \bibinfo{journal}{Phys. Rev.} \textbf{\bibinfo{volume}{B 69}},
  \bibinfo{pages}{115208} (\bibinfo{year}{2004}).

\bibitem[{\citenamefont{Schulthess et~al.}(2005)\citenamefont{Schulthess,
  Temmerman, Szotek, Butler, and Stocks}}]{Schulthess:2005_a}
\bibinfo{author}{\bibfnamefont{T.}~\bibnamefont{Schulthess}},
  \bibinfo{author}{\bibfnamefont{W.~M.} \bibnamefont{Temmerman}},
  \bibinfo{author}{\bibfnamefont{Z.}~\bibnamefont{Szotek}},
  \bibinfo{author}{\bibfnamefont{W.~H.} \bibnamefont{Butler}},
  \bibnamefont{and} \bibinfo{author}{\bibfnamefont{G.~M.}
  \bibnamefont{Stocks}}, \bibinfo{journal}{Nature Materials}
  \textbf{\bibinfo{volume}{4}}, \bibinfo{pages}{838} (\bibinfo{year}{2005}).

\bibitem[{\citenamefont{Achenbach and Schuster}(1981)}]{Achenbach:1981_a}
\bibinfo{author}{\bibfnamefont{G.}~\bibnamefont{Achenbach}} \bibnamefont{and}
  \bibinfo{author}{\bibfnamefont{H.~U.} \bibnamefont{Schuster}},
  \bibinfo{journal}{Z. anorg. allg. Chem.} \textbf{\bibinfo{volume}{475}},
  \bibinfo{pages}{9} (\bibinfo{year}{1981}).

\bibitem[{\citenamefont{Bronger et~al.}(1986)\citenamefont{Bronger,
  {M\"{u}ller}, {H\"{o}ppner}, and Schuster}}]{Bronger:1986_a}
\bibinfo{author}{\bibfnamefont{W.}~\bibnamefont{Bronger}},
  \bibinfo{author}{\bibfnamefont{P.}~\bibnamefont{{M\"{u}ller}}},
  \bibinfo{author}{\bibfnamefont{R.}~\bibnamefont{{H\"{o}ppner}}},
  \bibnamefont{and} \bibinfo{author}{\bibfnamefont{H.~U.}
  \bibnamefont{Schuster}}, \bibinfo{journal}{Z. anorg. allg. Chem.}
  \textbf{\bibinfo{volume}{539}}, \bibinfo{pages}{175} (\bibinfo{year}{1986}).

\bibitem[{\citenamefont{{Ma\v{s}ek} et~al.}(2004)\citenamefont{{Ma\v{s}ek},
  Turek, {Kudrnovsk\'{y}}, {M\'{a}ca}, and Drchal}}]{Masek:2004_a}
\bibinfo{author}{\bibfnamefont{J.}~\bibnamefont{{Ma\v{s}ek}}},
  \bibinfo{author}{\bibfnamefont{I.}~\bibnamefont{Turek}},
  \bibinfo{author}{\bibfnamefont{J.}~\bibnamefont{{Kudrnovsk\'{y}}}},
  \bibinfo{author}{\bibfnamefont{F.}~\bibnamefont{{M\'{a}ca}}},
  \bibnamefont{and} \bibinfo{author}{\bibfnamefont{V.}~\bibnamefont{Drchal}},
  \bibinfo{journal}{Acta Phys. Pol.} \textbf{\bibinfo{volume}{A 105}},
  \bibinfo{pages}{637} (\bibinfo{year}{2004}), \eprint{cond-mat/0406314}.

\bibitem[{\citenamefont{Jungwirth et~al.}(2005)\citenamefont{Jungwirth, Wang,
  {Ma\v{s}ek}, Edmonds, {K\"{o}nig}, Sinova, Polini, Goncharuk, MacDonald,
  Sawicki et~al.}}]{Jungwirth:2005_b}
\bibinfo{author}{\bibfnamefont{T.}~\bibnamefont{Jungwirth}},
  \bibinfo{author}{\bibfnamefont{K.~Y.} \bibnamefont{Wang}},
  \bibinfo{author}{\bibfnamefont{J.}~\bibnamefont{{Ma\v{s}ek}}},
  \bibinfo{author}{\bibfnamefont{K.~W.} \bibnamefont{Edmonds}},
  \bibinfo{author}{\bibfnamefont{J.}~\bibnamefont{{K\"{o}nig}}},
  \bibinfo{author}{\bibfnamefont{J.}~\bibnamefont{Sinova}},
  \bibinfo{author}{\bibfnamefont{M.}~\bibnamefont{Polini}},
  \bibinfo{author}{\bibfnamefont{N.~A.} \bibnamefont{Goncharuk}},
  \bibinfo{author}{\bibfnamefont{A.~H.} \bibnamefont{MacDonald}},
  \bibinfo{author}{\bibfnamefont{M.}~\bibnamefont{Sawicki}},
  \bibnamefont{et~al.}, \bibinfo{journal}{Phys. Rev.}
  \textbf{\bibinfo{volume}{B 72}}, \bibinfo{pages}{165204}
  (\bibinfo{year}{2005}).

\bibitem[{\citenamefont{Liechtenstein et~al.}(1987)\citenamefont{Liechtenstein,
  Katsnelson, Antropov, and Gubanov}}]{Liechtenstein:1987_a}
\bibinfo{author}{\bibfnamefont{A.~I.} \bibnamefont{Liechtenstein}},
  \bibinfo{author}{\bibfnamefont{M.~I.} \bibnamefont{Katsnelson}},
  \bibinfo{author}{\bibfnamefont{V.~P.} \bibnamefont{Antropov}},
  \bibnamefont{and} \bibinfo{author}{\bibfnamefont{V.~A.}
  \bibnamefont{Gubanov}}, \bibinfo{journal}{J. Magn. Magn. Mater.}
  \textbf{\bibinfo{volume}{67}}, \bibinfo{pages}{65} (\bibinfo{year}{1987}).

\bibitem[{\citenamefont{{Ma\v{s}ek}}(1991)}]{Masek:1991_a}
\bibinfo{author}{\bibfnamefont{J.}~\bibnamefont{{Ma\v{s}ek}}},
  \bibinfo{journal}{Solid State Commun.} \textbf{\bibinfo{volume}{78}},
  \bibinfo{pages}{351} (\bibinfo{year}{1991}).

\bibitem[{\citenamefont{Sandratskii and Bruno}(2002)}]{Sandratskii:2002_a}
\bibinfo{author}{\bibfnamefont{L.~M.} \bibnamefont{Sandratskii}}
  \bibnamefont{and} \bibinfo{author}{\bibfnamefont{P.}~\bibnamefont{Bruno}},
  \bibinfo{journal}{Phys. Rev.} \textbf{\bibinfo{volume}{B 66}},
  \bibinfo{pages}{134435} (\bibinfo{year}{2002}).

\bibitem[{\citenamefont{Bhattacharjee et~al.}(1983)\citenamefont{Bhattacharjee,
  Fishman, and Coqblin}}]{Bhattacharjee:1983_a}
\bibinfo{author}{\bibfnamefont{A.~K.} \bibnamefont{Bhattacharjee}},
  \bibinfo{author}{\bibfnamefont{G.}~\bibnamefont{Fishman}}, \bibnamefont{and}
  \bibinfo{author}{\bibfnamefont{B.}~\bibnamefont{Coqblin}},
  \bibinfo{journal}{Physica} \textbf{\bibinfo{volume}{B+C 117-118}},
  \bibinfo{pages}{449} (\bibinfo{year}{1983}).

\bibitem[{\citenamefont{Dietl}(1994)}]{Dietl:1994_a}
\bibinfo{author}{\bibfnamefont{T.}~\bibnamefont{Dietl}}, in
  \emph{\bibinfo{booktitle}{Handbook of Semiconductors}}, edited by
  \bibinfo{editor}{\bibfnamefont{S.}~\bibnamefont{Mahajan}}
  (\bibinfo{publisher}{North Holland, Amsterdam}, \bibinfo{year}{1994}),
  vol.~\bibinfo{volume}{3B}, p. \bibinfo{pages}{1251}.

\bibitem[{\citenamefont{Harrison}(1980)}]{Harrison:1980_a}
\bibinfo{author}{\bibfnamefont{W.}~\bibnamefont{Harrison}},
  \emph{\bibinfo{title}{Electronic Structure and the Properties of Solid}}
  (\bibinfo{publisher}{Freeman, San Francisco}, \bibinfo{year}{1980}).

\bibitem[{\citenamefont{Sato et~al.}(2004)\citenamefont{Sato, Schweika,
  Dederichs, and Katayama-Yoshida}}]{Sato:2004_a}
\bibinfo{author}{\bibfnamefont{K.}~\bibnamefont{Sato}},
  \bibinfo{author}{\bibfnamefont{W.}~\bibnamefont{Schweika}},
  \bibinfo{author}{\bibfnamefont{P.~H.} \bibnamefont{Dederichs}},
  \bibnamefont{and}
  \bibinfo{author}{\bibfnamefont{H.}~\bibnamefont{Katayama-Yoshida}},
  \bibinfo{journal}{Phys. Rev.} \textbf{\bibinfo{volume}{B 70}},
  \bibinfo{pages}{201202(R)} (\bibinfo{year}{2004}).

\bibitem[{\citenamefont{Bergqvist et~al.}(2004)\citenamefont{Bergqvist,
  Eriksson, {Kudrnovsk\'{y}}, Drchal, Korzhavyi, and Turek}}]{Bergqvist:2004_a}
\bibinfo{author}{\bibfnamefont{L.}~\bibnamefont{Bergqvist}},
  \bibinfo{author}{\bibfnamefont{O.}~\bibnamefont{Eriksson}},
  \bibinfo{author}{\bibfnamefont{J.}~\bibnamefont{{Kudrnovsk\'{y}}}},
  \bibinfo{author}{\bibfnamefont{V.}~\bibnamefont{Drchal}},
  \bibinfo{author}{\bibfnamefont{P.}~\bibnamefont{Korzhavyi}},
  \bibnamefont{and} \bibinfo{author}{\bibfnamefont{I.}~\bibnamefont{Turek}},
  \bibinfo{journal}{Phys. Rev. Lett.} \textbf{\bibinfo{volume}{93}},
  \bibinfo{pages}{137202} (\bibinfo{year}{2004}).

\bibitem[{\citenamefont{Bouzerar et~al.}(2005)\citenamefont{Bouzerar, Ziman,
  and {Kudrnovsk\'{y}}}}]{Bouzerar:2004_a}
\bibinfo{author}{\bibfnamefont{G.}~\bibnamefont{Bouzerar}},
  \bibinfo{author}{\bibfnamefont{T.}~\bibnamefont{Ziman}}, \bibnamefont{and}
  \bibinfo{author}{\bibfnamefont{J.}~\bibnamefont{{Kudrnovsk\'{y}}}},
  \bibinfo{journal}{Europhys. Lett.} \textbf{\bibinfo{volume}{69}},
  \bibinfo{pages}{812} (\bibinfo{year}{2005}).


\end{thebibliography}

\end{document}